\documentstyle[aps,floats,prl,preprint,psfig]{revtex}
\begin{document}
\author{
Santi Prestipino~$^*$ and Paolo V. Giaquinta~\footnote[1]{
Istituto Nazionale per la Fisica della Materia (INFM),
Unit\`a di Ricerca di Messina, Italy; and
Universit\`a degli Studi di Messina, Dipartimento di Fisica,
Contrada Papardo, 98166 Messina, Italy;
e-mail: {\tt Santi.Prestipino@unime.it} and {\tt Paolo.Giaquinta@unime.it}}
}
\title{The concavity of entropy \\ and extremum principles in thermodynamics}
\maketitle
\begin{abstract}
~~We revisit the concavity property of the thermodynamic entropy in
order to formulate a general proof of the minimum energy principle
as well as of other equivalent extremum principles that are valid
for thermodynamic potentials and corresponding Massieu functions
under different constraints. The current derivation aims at
providing a coherent formal framework for such principles which
may be also pedagogically useful as it fully exploits and
highlights the equivalence between different schemes. We also
elucidate the consequences of the extremum principles for the
general shape of thermodynamic potentials in relation to
first-order phase transitions.

\vspace{5mm}

\end{abstract}

\newpage
\section{Introduction}

In developing the formal structure of thermodynamics, one usually
starts from the maximum entropy principle as the guiding principle
that is used to predict the equilibrium conditions which apply to
isolated systems. This is, actually, the way the subject was
introduced by Callen in his celebrated book~\cite{Callen}.
Crucially important in his presentation of equilibrium
thermodynamics is the proof of the equivalence between different
representations which are based on different choices of the
natural variables that can be introduced in order to describe the
macroscopic state of the system. Such a proof implies an extension
of the extremum principle to other thermodynamic schemes. However,
the approach that is usually pursued to justify the above
equivalence, albeit physically well founded, may be not completely
satisfactory on the formal side. More specifically, the ordinary
proof of the minimum energy principle is formulated for a system
with just one degree of freedom and its extension to a
thermodynamic space of higher dimensionality is not, in our
opinion, straightforward.

This state of affairs is also probably responsible for the
somewhat uncertain status of ``thermodynamic potentials'' with
multiple minima that are usually introduced when discussing
first-order phase transitions~\cite{Callen}. Actually, it is not
immediately clear how such a potential, which fails to fulfil the
convexity requirement, is related to the fundamental equation of
the system, unless one explicitly intends to represent a Landau
free energy, {\it i.e.}, the outcome of a mean-field calculation.

For the above reasons, we believe that it can be useful to revisit
the proof of the extremum principles used in thermodynamics in
order to place all representations on a more clear mathematical
basis which may turn useful also for a pedagogical presentation of
the subject.

The outline of the paper is as follows: in Section 2, we provide a
general proof of the minimum energy principle based on the
concavity property of the entropy function. Then, upon discussing
the case of a system in contact with a reservoir, we derive in
Section 3 other forms of the extremum principle which apply to the
generalized thermodynamic potentials and related Massieu
functions. We also analyze the pattern of singularities of, say,
the Gibbs free energy in proximity to a first-order transition
point. Some further remarks and a brief summary of the main
results are given in the Conclusions.

\section{The minimum energy principle}
Following Callen~\cite{Callen}, the fundamental problem of
thermodynamics is to find the equilibrium state of an {\it overall
isolated} macroscopic system following the removal of one or more
internal constraints, {\it i.e.}, walls restrictive with respect
to the exchange of energy and, possibly, of other extensive
quantities between the various parts of the system. As is well
known, the solution to this problem can be cast in terms of the
maximum entropy principle: the equilibrium state eventually
singled out by the system is the one that maximizes the total
entropy out of the variety of states that are compatible with the
residual constraints.

Thermodynamics essentially postulates three properties for the
entropy $S$~\cite{Callen}: 1) $S$ is a well-behaved, first-order
homogeneous function of the extensive parameters (a property
leading to Euler's theorem (\ref{eq1})); 2) $S$ is additive over
disjoint subsystems; and 3) the partial derivative of $S$ with
respect to the energy $U$ is strictly positive (implying that the
temperature $T>0$). In particular, this latter condition allows
one to express the energy as a function of the entropy as well as
of the other extensive parameters $X_i$ which specify the state of
the system, in such a way that the knowledge of
$U(S,X_1,X_2,\ldots )$ is equivalent to that of
$S(U,X_1,X_2,\ldots )$. In the following, we shall ignore any
exception to the above conditions such as those arising from the
existence of long-ranged interactions between the constituent
particles (additivity and, possibly, extensivity violated), or
from an effective ergodicity breaking (which causes the
unattainability of equilibrium).

In order to set the stage for our subsequent reasoning, we
consider an isolated system described by the energy $U$, the
volume $V$, and the number of particles $N$ as the only extensive
parameters. For this system, the entropy is written as
\begin{equation}
S(U,V,N)=S_U(U,V,N)U+S_V(U,V,N)V+S_N(U,V,N)N\,,
\label{eq1}
\end{equation}
where $S_X(\cdots )$ is the partial derivative of $S$ with respect
to $X$ and $S_U(U,V,N)>0$. This condition allows one to solve
Eq.\,(\ref{eq1}) in $U$,
\begin{equation}
U(S,V,N)=T(S,V,N)S-P(S,V,N)V+\mu (S,V,N)N\,,
\label{eq2}
\end{equation}
where $T(S,V,N)\equiv U_S(S,V,N)$ is the temperature,
$P(S,V,N)\equiv -U_V(S,V,N)$ is the pressure, and
$\mu (S,V,N)\equiv U_N(S,V,N)$ is the chemical potential.

It is useful to recall that, given a function $y=Y(x,\alpha )$ with
$Y_x(x,\alpha )\neq 0$, the variable $x$ can be expressed, on
fairly general grounds, in terms of $y$ as $x=X(y,\alpha )$, with
\begin{equation}
X_y(y,\alpha )=\frac{1}{Y_x(X(y,\alpha ),\alpha)}\,\,\,\,{\rm
and}\,\,\,\, X_{\alpha}(y,\alpha )=-\frac{Y_{\alpha}(X(y,\alpha
),\alpha)}{Y_x(X(y,\alpha ),\alpha)}\,.
\label{eq3}
\end{equation}
Then, the partial derivatives of $S$ can be identified as:
\begin{eqnarray}
S_U(U,V,N) &=& \frac{1}{T(S(U,V,N),V,N)}\equiv\left( \frac{1}{T}\right)
(U,V,N)\,;
\nonumber \\
S_V(U,V,N) &=& \frac{P(S(U,V,N),V,N)}{T(S(U,V,N),V,N)}\equiv\left(
\frac{P}{T}\right) (U,V,N)\,;
\nonumber \\
S_N(U,V,N) &=& -\frac{\mu (S(U,V,N),V,N)}{T(S(U,V,N),V,N)}\equiv -\left(
\frac{\mu}{T}\right) (U,V,N)\,.
\label{eq4}
\end{eqnarray}

It is now possible to show that the maximum principle, along with
the extensivity and additivity properties, underlies the concavity
of the entropy function. Let $\lambda$ be any number with
$0<\lambda<1$. Furthermore, let $(U_1,V_1,N_1)$ and
$(U_2,V_2,N_2)$ identify two generic macroscopic states of the
system. Imagine, then, to form a single isolated system by putting
together a fraction $1-\lambda$ of $(U_1,V_1,N_1)$ and a fraction
$\lambda$ of $(U_2,V_2,N_2)$. Once the exchange of $U,V$, and $N$
between the two subsystems is allowed, the overall system evolves
until its entropy reaches a value $S((1-\lambda)U_1+\lambda
U_2,\ldots )$ which is larger than (or at most equal to) the
initial value $S((1-\lambda)U_1,\ldots )+S(\lambda U_2,\ldots )$.
Thanks to the extensivity of the entropy, this readily implies
that $S$ is a {\it concave} function of $U,V$, and $N$:
\begin{equation}
S((1-\lambda)U_1+\lambda U_2,\ldots )\geq (1-\lambda)S(U_1,V_1,N_1)+
\lambda S(U_2,V_2,N_2)\,.
\label{eq5}
\end{equation}
In deriving Eq.\,(\ref{eq5}), we have tacitly assumed that the
state space is a convex set. This assumption is physically
reasonable as, for instance, the set $U>U_0,V>0$, and $N>0$ is an
open convex subset of ${\cal R}^3$~\cite{url}.

As is well known, the occurrence of an equality sign in
(\ref{eq5}) is linked with the phenomenon of phase coexistence,
{\it i.e.}, with the occurrence of a first-order phase transition.
In such a case, the system is macroscopically inhomogeneous and
the state $((1-\lambda)U_1+\lambda U_2,\ldots )$ is interpreted as
a mixture of the phases $(U_1,V_1,N_1)$ and $(U_2,V_2,N_2)$.
Hence, unless two distinct thermodynamic phases can coexist,
$S(U,V,N)$ is a {\it strictly} concave function of $U,V,N$ ({\it
i.e.}, Eq.\,(\ref{eq5}) holds as a strict inequality).

Owing to the concavity of the entropy, the Hessian form of $S$ is
negative semidefinite (see Theorem 3 in Appendix A of \cite{url}),
a property that is usually expressed as ${\rm d}^2S\leq 0$. Upon taking in
Eq.\,(\ref{eq5}) $N_1=N_2$, the $(U,V)$ Hessian of $S$ turns out to
be negative semidefinite as well. This condition yields the
inequalities~\cite{note}:
\begin{equation}
S_{UU}\leq 0\,,\,\,S_{VV}\leq 0\,,\,\,{\rm
and}\,\,\,S_{UU}S_{VV}-S_{UV}^2\geq 0\,,
\label{eq6}
\end{equation}
which are to be satisfied for all $(U,V,N)$. These inequalities
represent the conditions of thermodynamic stability for a system
at equilibrium. It follows from Eqs.\,(\ref{eq6}) that the
constant-volume and constant-pressure heat capacities and the
isothermal and isentropic compressibilities are non negative
quantities~\cite{Callen}. Similar stability conditions do also
hold for $U(S,V,N)$. In fact, a rather straightforward calculation
along the same lines as those leading to
Eqs.\,(B33)-(B35) of \cite{url}, yields:
\begin{eqnarray}
U_{SS}(S,V,N)&=&T_S(S,V,N)=-T(S,V,N)^3S_{UU}(U(S,V,N),V,N)\geq 0\,;
\nonumber \\
U_{VV}(S,V,N)&=&-P_V(S,V,N)=-T(S,V,N)
\nonumber \\
&\times&\left[ S_{VV}(U(S,V,N),V,N)-2S_{UV}(U(S,V,N),V,N)\cdot
P(S,V,N)\right.
\nonumber \\
&+&\left. S_{UU}(U(S,V,N),V,N)\cdot P(S,V,N)^2\right] \geq 0\,;
\nonumber \\
U_{SS}U_{VV}-U_{SV}^2&=&T(S,V,N)^4\left( S_{UU}S_{VV}-S_{UV}^2\right) \geq 0\,.
\label{eq7}
\end{eqnarray}
As for the sign of $U_{VV}$, note that the quantity within square
brackets is the value taken in $(1,-P(S,V,N))$ by the $(U,V)$
Hessian of $S$ relative to $(U(S,V,N),V,N)$.

Now, let us consider an isolated system composed of two weakly
interacting subsystems (say, 1 and 2), not necessarily made of the
same substance, which, after removing an internal wall, may
exchange energy and one more extensive quantity ({\it e.g.}, the
volume $V$) between each other. Hereafter, we shall omit in the
notation explicit reference to any other extensive parameter that
is separately conserved for each subsystem. The equilibrium state
eventually reached by the system is the state that maximizes the
total entropy
\begin{equation}
\tilde{S}(U_1,V_1;U,V)=S^{(1)}(U_1,V_1)+S^{(2)}(U-U_1,V-V_1)
\label{eq8}
\end{equation}
with respect to the parameters of subsystem 1. In Eq.\,(\ref{eq8}),
$U$ and $V$ are the (fixed) values of energy and volume pertaining
to the entire system. It is worth noting that the concavity of $S^{(1)}$
and $S^{(2)}$ implies that $\tilde{S}$ as well is a concave function
of $U_1$ and $V_1$:
\begin{eqnarray}
&& \tilde{S}((1-\lambda )U_1^{(A)}+\lambda U_1^{(B)},
(1-\lambda )V_1^{(A)}+\lambda V_1^{(B)};U,V)
\nonumber \\
&=& S^{(1)}((1-\lambda )U_1^{(A)}+\lambda U_1^{(B)},\ldots )+
S^{(2)}(U-(1-\lambda )U_1^{(A)}-\lambda U_1^{(B)},\ldots )
\nonumber \\
&=& S^{(1)}((1-\lambda )U_1^{(A)}+\lambda U_1^{(B)},\ldots )+
S^{(2)}((1-\lambda )(U-U_1^{(A)})+\lambda (U-U_1^{(B)}),\ldots )
\nonumber \\
&\geq& (1-\lambda )\tilde{S}(U_1^{(A)},V_1^{(A)};U,V)+
\lambda\tilde{S}(U_1^{(B)},V_1^{(B)};U,V)\,.
\label{eq9}
\end{eqnarray}
As a result, the $(U_1,V_1)$ Hessian of $\tilde{S}$ is negative
semidefinite, which implies:
\begin{equation}
\tilde{S}_{U_1U_1}\leq 0\,,\,\,\tilde{S}_{V_1V_1}\leq 0\,,\,\,{\rm
and}\,\,\,\tilde{S}_{U_1U_1}\tilde{S}_{V_1V_1}-\tilde{S}_{U_1V_1}^2\geq
0\,. \label{eq10}
\end{equation}

We now turn to the maximum condition for $\tilde{S}$. The
necessary conditions for any extremal point $(U_1^0,V_1^0)$ of
$\tilde{S}$ are:
\begin{equation}
\tilde{S}_{U_1}=0\,\,\,\,\Longrightarrow\,\,\,\,\left( \frac{1}{T}\right)
^{(1)}(U_1^0,V_1^0)=\left( \frac{1}{T}\right) ^{(2)}(U-U_1^0,V-V_1^0)\,;
\label{eq11}
\end{equation}
\begin{equation}
\tilde{S}_{V_1}=0\,\,\,\,\Longrightarrow\,\,\,\,\left( \frac{P}{T}\right)
^{(1)}(U_1^0,V_1^0)=\left( \frac{P}{T}\right) ^{(2)}(U-U_1^0,V-V_1^0)\,.
\label{eq12}
\end{equation}
These equations merely express the well known fact that the
conditions of thermal and mechanical equilibrium between
subsystems 1 and 2 entail the same values of temperature and
pressure for both subsystems. Furthermore, because of the
concavity property, any extremum of $\tilde{S}$ is necessarily a
global maximum, which is moreover strict if $\tilde{S}$ is
strictly concave (see Theorem 1 and the corollary of Theorem 2 in
Appendix A of \cite{url}).

The solution to Eqs.\,(\ref{eq11}) and (\ref{eq12}) is generally
unique (say, $U_1^0(U,V)$ and $V_1^0(U,V)$). In fact, even if the
final equilibrium state hosted two coexisting phases, the values
of energy and volume of each subsystem would be uniquely
determined from (\ref{eq11}) and (\ref{eq12}), owing to the fact
that $N_1$ and $N_2$ are fixed. Note that the {\it derivatives} of
\begin{equation}
{\mathcal{S}}(U,V)\equiv\tilde{S}(U_1^0(U,V),V_1^0(U,V);U,V)
\label{eq13}
\end{equation}
are well-defined only when the point of maximum of $\tilde{S}$ is
unique. In this case:
\begin{eqnarray}
{\mathcal{S}}_U &=& \tilde{S}_{U_1}\cdot (U_1^0)_U+
\tilde{S}_{V_1}\cdot (V_1^0)_U+\tilde{S}_U=\tilde{S}_U
\nonumber \\
&=& \left( \frac{1}{T}\right) ^{(2)}(U-U_1^0(U,V),V-V_1^0(U,V))
\equiv\left( \frac{1}{T}\right) (U,V)>0\,;
\nonumber \\
{\mathcal{S}}_V &=& \tilde{S}_{U_1}\cdot (U_1^0)_V+
\tilde{S}_{V_1}\cdot (V_1^0)_V+\tilde{S}_V=\tilde{S}_V
\nonumber \\
&=& \left( \frac{P}{T}\right) ^{(2)}(U-U_1^0(U,V),V-V_1^0(U,V))
\equiv\left( \frac{P}{T}\right) (U,V)\,.
\label{eq14}
\end{eqnarray}
We further notice that, if $S^{(1)}$ and $S^{(2)}$ happen to be
the same function $S$ ({\it i.e.}, they pertain to the same
substance), then ${\mathcal{S}}(U,V)=S(U,V,N_1+N_2)$ is a concave
function of $U$ and $V$.

We now proceed to demonstrate that the maximum entropy principle
can be reformulated as a minimum principle for the total energy,
under a constraint on the value of the total entropy. To begin
with, we call $U^{(1)}(S_1,V_1)$ the energy function of subsystem
1, obtained from $S_1=S^{(1)}(U_1,V_1)$ by solving the latter with
respect to $U_1$. Similarly, let $U^{(2)}(S_2,V_2)$ be the energy
of subsystem 2. The crucial step in our proof of the minimum
energy principle will be to show that $U^{(1)}$ and $U^{(2)}$ are
convex functions. To this aim, all we need to recall is that
$U^{(1)}(S,V)$ (as well as $U^{(2)}$) is an increasing function of
its former argument, since $U^{(1)}_{S}(S,V)=T^{(1)}(S,V)>0$.
Setting $S_A=S^{(1)}(U_A,V_A)$ and $S_B=S^{(1)}(U_B,V_B)$, the
concavity of $S^{(1)}$,
\begin{equation}
S^{(1)}((1-\lambda )U_A+\lambda U_B,(1-\lambda )V_A+\lambda V_B)\geq
(1-\lambda )S^{(1)}(U_A,V_A)+\lambda S^{(1)}(U_B,V_B)\,,
\label{eq15}
\end{equation}
can be rewritten as
\begin{equation}
(1-\lambda )S_A+\lambda S_B\leq
S^{(1)}((1-\lambda )U^{(1)}(S_A,V_A)+\lambda U^{(1)}(S_B,V_B),
(1-\lambda )V_A+\lambda V_B)\,.
\label{eq16}
\end{equation}
After evaluating $U^{(1)}$ at the point $((1-\lambda )S_A+\lambda
S_B,(1-\lambda )V_A+\lambda V_B)$, it immediately follows from
Eq.\,(\ref{eq16}) that:
\begin{equation}
U^{(1)}((1-\lambda )S_A+\lambda S_B,(1-\lambda )V_A+\lambda V_B)\leq
(1-\lambda )U^{(1)}(S_A,V_A)+\lambda U^{(1)}(S_B,V_B)\,.
\label{eq17}
\end{equation}
Under such premises, we shall now prove that, if the total entropy
takes the value $S\equiv {\mathcal{S}}(U,V)$, then the function
\begin{equation}
\tilde{U}(S_1,V_1;S,V)=U^{(1)}(S_1,V_1)+U^{(2)}(S-S_1,V-V_1)\,,
\label{eq18}
\end{equation}
attains its minimum for $V_1=V_1^0(U,V)$ and
$S_1=S^{(1)}(U_1^0(U,V),V_1^0(U,V))\equiv S_1^0(U,V)$, where
$(U_1^0(U,V),V_1^0(U,V))$ is {\it any} solution to
Eqs.\,(\ref{eq11}) and (\ref{eq12}). Moreover, the minimum value
of $\tilde{U}$ is $U$.

We start noting, using an argument identical to that already
developed for $\tilde{S}$, that $\tilde{U}$ is a convex function
of $S_1$ and $V_1$. Therefore, in order to achieve our goal, all
we need to show is that the first-order derivatives of $\tilde{U}$
at $(S_1^0(U,V),V_1^0(U,V))$ are both zero, since then the
convexity of $\tilde{U}$ allows one to conclude that the extremum
is a global minimum.

The general expression of the first-order derivatives of
$\tilde{U}$ is:
\begin{eqnarray}
\tilde{U}_{S_1} &=& T^{(1)}(S_1,V_1)-T^{(2)}(S-S_1,V-V_1)
\nonumber \\
&=& S^{(1)}_{U_1}(U^{(1)}(S_1,V_1),V_1)^{-1}-
S^{(2)}_{U_2}(U^{(2)}(S-S_1,V-V_1),V-V_1)^{-1}
\nonumber \\
&\propto& \left( \frac{1}{T}\right) ^{(2)}(U^{(2)}(S-S_1,V-V_1),V-V_1)-
\left( \frac{1}{T}\right) ^{(1)}(U^{(1)}(S_1,V_1),V_1)\,;
\nonumber \\
\label{eq19}
\end{eqnarray}
\begin{eqnarray}
\tilde{U}_{V_1} &=& -P^{(1)}(S_1,V_1)+P^{(2)}(S-S_1,V-V_1)
\nonumber \\
&=& \left( \frac{P}{T}\right) ^{(2)}(U^{(2)}(S-S_1,V-V_1),V-V_1)
\cdot T^{(2)}(S-S_1,V-V_1)
\nonumber \\
&-& \left( \frac{P}{T}\right) ^{(1)}(U^{(1)}(S_1,V_1),V_1)
\cdot T^{(1)}(S_1,V_1)\,.
\label{eq20}
\end{eqnarray}
When $S_1=S_1^0(U,V)$ and $V_1=V_1^0(U,V)$, the energy of subsystem 1 is
\begin{equation}
U^{(1)}(S_1^0(U,V),V_1^0(U,V))=U_1^0(U,V)\,,
\label{eq21}
\end{equation}
since $U^{(1)}$ is the inverse of $S^{(1)}$. Moreover,
\begin{eqnarray}
S^{(2)}(U-U_1^0(U,V),V-V_1^0(U,V)) &=& {\mathcal{S}}(U,V)-
S^{(1)}(U_1^0(U,V),V_1^0(U,V))
\nonumber \\
&=& S-S_1^0(U,V)\,,
\label{eq22}
\end{eqnarray}
which implies:
\begin{equation}
U^{(2)}(S-S_1^0(U,V),V-V_1^0(U,V))=U-U_1^0(U,V)\,.
\label{eq23}
\end{equation}
Given Eqs.\,(\ref{eq21}) and (\ref{eq23}), it follows from
Eq.\,(\ref{eq11}) that $\tilde{U}_{S_1}=0$. This result, when
combined with Eq.\,(\ref{eq12}), yields $\tilde{U}_{V_1}=0$.
Finally, the absolute minimum of $\tilde{U}$ is clearly $U$ (see
Eqs.\,(\ref{eq21}) and (\ref{eq23})) and this completes our proof.
We further note that
${\mathcal{U}}(S,V)=\min_{S_1,V_1}\tilde{U}(S_1,V_1;S,V)$ is the
inverse function of ${\mathcal{S}}(U,V)$. In fact, for arbitrary
$U$ and $V$, we have shown that
${\mathcal{U}}({\mathcal{S}}(U,V),V)=U$. We point out that the
hypothesis according to which the subsystems can only exchange two
extensive parameters between each other does not affect the
generality of our proof of the minimum energy principle; rather,
this restriction simply avoids the use of a cumbersome notation.

A different derivation of the minimum energy principle, which does
not resort to the convexity of $\tilde{U}$, is also viable. In
this case, one must show that the $(S_1,V_1)$ Hessian of $\tilde{U}$
for $V_1=V_1^0(U,V)$ and $S_1=S_1^0(U,V)$ is positive definite.
Actually, what can be achieved this way is a weaker result, {\it
i.e.}, that the Hessian of $\tilde{U}$ at the extremal point is
positive {\it semi}definite.

Using Eqs.\,(\ref{eq7}), one immediately gets for $V_1=V_1^0(U,V)$
and $S_1=S_1^0(U,V)$:
\begin{eqnarray}
\tilde{U}_{S_1S_1} &=& -T^0(U,V)^3\tilde{S}_{U_1U_1}\geq 0\,;
\nonumber \\
\tilde{U}_{V_1V_1} &=& -T^0(U,V)\left( \tilde{S}_{V_1V_1}-
2\tilde{S}_{U_1V_1}P^0(U,V)+
\tilde{S}_{U_1U_1}P^0(U,V)^2\right) \geq 0\,;
\nonumber \\
\tilde{U}_{S_1V_1} &=& -T^0(U,V)^2\left(
-\tilde{S}_{U_1U_1}P^0(U,V)+\tilde{S}_{U_1V_1}\right) \,,
\label{eq24}
\end{eqnarray}
where $T^0(U,V)=T^{(1)}(S_1^0(U,V),V_1^0(U,V))$,
$P^0(U,V)=P^{(1)}(S_1^0(U,V),V_1^0(U,V))$, whereas the arguments
of the second-order $\tilde{S}$ derivatives are
$U_1^0(U,V),V_1^0(U,V),U$, and $V$. Equations (\ref{eq24}) also lead to:
\begin{equation}
\tilde{U}_{S_1S_1}\tilde{U}_{V_1V_1}-\tilde{U}_{S_1V_1}^2=
T^0(U,V)^4\left( \tilde{S}_{U_1U_1}\tilde{S}_{V_1V_1}-
\tilde{S}_{U_1V_1}^2\right) \geq 0\,,
\label{eq25}
\end{equation}
which concludes the proof that the Hessian of $\tilde{U}$ is
positive semidefinite in the final equilibrium state. Besides the
general impossibility to conclude, on account of the above
inequalities, that the $\tilde{U}$ extremum is a minimum (in fact,
we are abstaining from using the convexity of $\tilde{U}$), the
intrinsic limitation of the latter proof of the minimum energy
principle lies in the fact that it only applies when the
subsystems are allowed to mutually exchange at most {\it two}
extensive parameters. In fact, only in this case the character of
the Hessian of $\tilde{U}$ in the final equilibrium state can be
decided in a relatively simple way on the basis of the sign of the
second-order derivatives.

\section{Minimum principles for \\ other thermodynamic potentials}

Thermodynamic representations other than the entropy or the energy
schemes arise when describing the equilibrium of a system that is
in contact with a reservoir. Let us consider, for instance, an
energy reservoir (heat bath). By definition, the temperature of a
heat bath is the same in any state, {\it i.e.},
\begin{equation}
(S_r)_{U_r}(U_r,V_r)=\frac{1}{T}\,,
\label{b1}
\end{equation}
a constant number which does not depend on the energy $U_r$ or the
volume $V_r$ of the reservoir. Hence, the entropy of a heat bath
reads as
\begin{equation}
S_r(U_r,V_r)=\frac{U_r}{T}+f(V_r)\,,
\label{b2}
\end{equation}
where $f$ is an unspecified, concave function of $V_r$. As usual,
we omitted to specify the particle number in the notation.

When a system with an entropy function $S(U,V)$ is brought into
contact with a heat bath, the joint system being isolated from the
outside environment, the final equilibrium state maximizes the
total entropy
\begin{equation}
\tilde{S}(U;U_{\rm tot},V,V_r)=S(U,V)+S_r(U_{\rm tot}-U,V_r)\,,
\label{b3}
\end{equation}
for fixed $U_{\rm tot}=U+U_r,V$, and $V_r$ (we assume that a rigid
and impermeable wall keeps the system separate from the bath). The
maximum condition then reads:
\begin{equation}
\tilde{S}_U=0\,\,\,\,\Longrightarrow\,\,\,\,S_U(U,V)=\frac{1}{T}\,,
\label{b4}
\end{equation}
which is equivalent to $U_S(S,V)=T$, $U(S,V)$ being the inverse
function of $S(U,V)$. It might happen that the solution $U^0$ to
Eq.\,(\ref{b4}) is not unique. However, if $S(U,V)$ is strictly
concave, there is a unique point of maximum $U^0(T,V)$ for
$\tilde{S}$, which represents the equilibrium value of the system
energy. In this case, the system entropy in the joint equilibrium
state is also well-defined, being $S(U^0(T,V),V)\equiv S^0(T,V)$
and $U(S^0(T,V),V)=U^0(T,V)$.

Let us now introduce the convex function of $U$ and $V$ given by~\cite{note2}
\begin{equation}
\tilde{F}(U;T,V)=U-TS(U,V)\,.
\label{b5}
\end{equation}
By simply looking at its derivatives,
\begin{equation}
\tilde{F}_U=1-TS_U\,\,\,\,{\rm and}\,\,\,\,\tilde{F}_{UU}=-TS_{UU}\,,
\label{b6}
\end{equation}
it is immediately apparent that the maximum condition for
$\tilde{S}$ is also the minimum condition for $\tilde{F}$. We call
$\tilde{F}$ a generalized thermodynamic potential. The minimum
value $F(T,V)$ of $\tilde{F}$ is the usual Helmholtz free energy.
In fact, $F(T,V)$ is the Legendre transform of $U(S,V)$ with
respect to $S$:
\begin{eqnarray}
F(T,V) &=& U^0(T,V)-TS(U^0(T,V),V)=U(S^0(T,V),V)-TS^0(T,V)
\nonumber \\
&=& \left[ U(S,V)-TS\right] _{S=S^0(T,V)}\,,
\label{b7}
\end{eqnarray}
where we observe that $S^0(T,V)$ is the unique solution to $U_S(S,V)=T$
(a more general case is treated below).
It is rather simple to calculate the first-order $F$ derivatives:
\begin{eqnarray}
F_T &=& U^0_T(T,V)-S(U^0(T,V),V)-TS_U(U^0(T,V),V)\cdot U^0_T(T,V)
\nonumber \\
&=& -S(U^0(T,V),V)=-S^0(T,V)\,;
\label{b8}
\end{eqnarray}
\begin{eqnarray}
F_V &=& U^0_V(T,V)-T\left[ S_U(U^0(T,V),V)\cdot U^0_V(T,V)+
S_V(U^0(T,V),V)\right]
\nonumber \\
&=& -TS_V(U^0(T,V),V)=-\frac{S_V(U^0(T,V),V)}{S_U(U^0(T,V),V)}
=U_V(S^0(T,V),V)
\nonumber \\
&=& -P(S^0(T,V),V)\equiv -P^0(T,V)\,.
\label{b9}
\end{eqnarray}

In order to calculate the second-order derivatives of $F$ we make
use of Eqs.\,(B5) and (B7) of \cite{url}:
\begin{eqnarray}
F_{TT} &=& -S^0_T(T,V)=-\frac{1}{U_{SS}(S^0(T,V),V)}<0\,;
\nonumber \\
F_{VV} &=& -P^0_V(T,V)=\frac{U_{SS}U_{VV}-U_{SV}^2}{U_{SS}(S^0(T,V),V)}
\geq 0\,.
\label{b10}
\end{eqnarray}
It thus follows that $F$ is a concave function of $T$ and a convex
function of $V$.

Let us now consider the case of multiple solutions to
Eq.\,(\ref{b4}). For instance, it may happen for a particular
value $T_c$ of $T$ that Eq.\,(\ref{b4}) is solved by all
$U\in\left[ U_A^0,U_B^0\right]$, with $U_A^0=U(S_A^0,V)$ and
$U_B^0=U(S_B^0,V)$. This occurs if, between $U_A^0$ and $U_B^0$,
$S(U,V)$ is a linear function of $U$ (first-order transition at
temperature $T_c$). In this case, $U_S(S,V)=T_c$
is satisfied for all $S\in\left[ S_A^0,S_B^0\right]$, and $U_S$
cannot be inverted as a function of $S$. However, the function
$\tilde{F}$ is still well-defined, along with its global minimum
$F(T_c,V)$. Furthermore, Eq.\,(\ref{b7}) still holds, provided we
call $S^0(T,V)$ the {\it unique} solution to $U(S,V)=U^0(T,V)$. In
particular,
\begin{equation}
\lim_{T\rightarrow T_C^-}F(T,V)=U_A^0-T_cS_A^0=U_B^0-T_cS_B^0=
\lim_{T\rightarrow T_C^+}F(T,V)\,,
\label{b11}
\end{equation}
which means that $F(T,V)$ is continuous for $T=T_c$. However,
$F(T,V)$ has a cusp-like singularity for $T=T_c$:
\begin{equation}
\lim_{T\rightarrow T_C^-}F_T(T,V)=-S_A^0\neq -S_B^0=
\lim_{T\rightarrow T_C^+}F_T(T,V)\,.
\label{b12}
\end{equation}

As a further example, let us consider the case of a system exchanging
energy and volume with a reservoir.
The values of temperature and pressure are both fixed for the reservoir:
\begin{equation}
(S_r)_{U_r}(U_r,V_r)=\frac{1}{T}\,\,\,\,{\rm and}\,\,\,\,
(S_r)_{V_r}(U_r,V_r)=\frac{P}{T}\,.
\label{b13}
\end{equation}
Therefore, the bath entropy is now fully specified as
\begin{equation}
S_r(U_r,V_r)=\frac{U_r+PV_r}{T}\,.
\label{b14}
\end{equation}
The maximum conditions for the total entropy
\begin{equation}
\tilde{S}(U,V;U_{\rm tot},V_{\rm tot})=
S(U,V)+S_r(U_{\rm tot}-U,V_{\rm tot}-V)
\label{b15}
\end{equation}
then read:
\begin{equation}
\tilde{S}_U=0\,\,\,\,\Longrightarrow\,\,\,\,S_U(U,V)=\frac{1}{T}\,;
\label{b16}
\end{equation}
\begin{equation}
\tilde{S}_V=0\,\,\,\,\Longrightarrow\,\,\,\,S_V(U,V)=\frac{P}{T}\,.
\label{b17}
\end{equation}
Assuming a unique solution for the above equations, one finds
$U=U^0(T,P)$ and $V=V^0(T,P)$. Upon assuming $S^0(T,P)$ to be
$S(U^0(T,P),V^0(T,P))$, one has $U(S^0(T,P),V^0(T,P))=U^0(T,P)$.
Moreover, Eqs.\,(\ref{b16}) and (\ref{b17}) are equivalent to
$U_S(S,V)=T$ and $U_V(S,V)=-P$.

As we did previously for a system in contact with a heat bath, it
is appropriate to introduce the auxiliary, convex function of $U$
and $V$, given by
\begin{equation}
\tilde{G}(U,V;T,P)=U-TS(U,V)+PV\,.
\label{b18}
\end{equation}
Clearly, $\tilde{G}$ attains its minimum for $U=U^0(T,P)$ and
$V=V^0(T,P)$, since
\begin{equation}
\tilde{G}_U=1-TS_U\,\,\,\,{\rm and}\,\,\,\,\tilde{G}_V=-TS_V+P
\label{b19}
\end{equation}
are both zero.
The minimum value $G(T,P)$ of $\tilde{G}$ is the Gibbs free energy.
In fact,
\begin{eqnarray}
G(T,P) &=& U^0(T,P)-TS(U^0(T,P),V^0(T,P))+PV^0(T,P)
\nonumber \\
&=& U(S^0(T,P),V^0(T,P))-TS^0(T,P)+PV^0(T,P)
\nonumber \\
&=& \left[ U(S,V)-TS+PV\right] _{S=S^0(T,P),V=V^0(T,P)}
\label{b20}
\end{eqnarray}
is the Legendre transform of $U(S,V)$ with respect to $S$ and $V$,
which have been replaced by their conjugate variables $T$ and $-P$
(observe that $(S^0(T,P),V^0(T,P))$ is, by our previous
assumption, the unique solution to $U_S(S,V)=T$ and
$U_V(S,V)=-P$). The first-order derivatives of $G$ are simply
calculated as:
\begin{eqnarray}
G_T &=& U^0_T(T,P)-S^0(T,P)-T\left[ S_U(U^0(T,P),V^0(T,P))\cdot
U^0_T(T,P)\right.
\nonumber \\
&+& \left. S_V(U^0(T,P),V^0(T,P))\cdot V^0_T(T,P)\right]
+PV^0_T(T,P)
\nonumber \\
&=& -S^0(T,P)\,;
\label{b21}
\end{eqnarray}
\begin{eqnarray}
G_P &=& U^0_P(T,P)-T\left[ S_U(U^0(T,P),V^0(T,P))\cdot
U^0_P(T,P)\right.
\nonumber \\
&+& \left. S_V(U^0(T,P),V^0(T,P))\cdot V^0_P(T,P)\right] +V^0(T,P)
+PV^0_P(T,P)
\nonumber \\
&=& V^0(T,P)\,,
\label{b22}
\end{eqnarray}
whereas, with the help of Eqs.\,(B20)-(B22) and
(B24) of \cite{url}, the second-order $G$ derivatives turn out to be:
\begin{eqnarray}
G_{TT} &=& -S^0_T(T,P)=-\frac{U_{VV}(S^0(T,P),V^0(T,P))}
{U_{SS}U_{VV}-U_{SV}^2}\leq 0\,;
\nonumber \\
G_{PP} &=& V^0_P(T,P)=-\frac{U_{SS}(S^0(T,P),V^0(T,P))}
{U_{SS}U_{VV}-U_{SV}^2}\leq 0\,;
\nonumber \\
G_{TT}G_{PP}-G_{TP}^2 &=& \left( U_{SS}U_{VV}-U_{SV}^2\right) ^{-1}>0\,.
\label{b23}
\end{eqnarray}
Hence, $G$ is a concave function of both $T$ and $P$.

Summing up, when a system is in thermal and mechanical contact
with a reservoir, it is $\tilde{G}$ (the generalized Gibbs
potential) that is minimum at equilibrium, not the Gibbs free
energy as is sometimes stated. Similarly, if the wall between the
system and the reservoir is permeable to the flow of energy and
particles while being restrictive to volume, it is
$\tilde{A}(U,N;T,V,\mu )=U-TS(U,V,N)-\mu N$ that is minimized at
equilibrium, its minimum value being the system grand potential.

Needless to say, equivalent maximum principles hold for the
functions $S(U,V)-(1/T)U$ and $S(U,V)-(1/T)U-(P/T)V$, equal to
$-(1/T)\tilde{F}(U;T,V)$ and $-(1/T)\tilde{G}(U,V;T,P)$,
respectively. Their maximum loci correspond to the usual
Massieu functions.

It is worth observing that the minimization of the generalized
Gibbs potential is correct also in the common situation of
determining the equilibrium between two systems that are in
contact with the same energy and volume reservoir. In fact,
the maximum condition for the total entropy
\begin{equation}
\tilde{S}=S^{(1)}(U_1,V_1)+S^{(2)}(U_2,V_2)+
\frac{U_{\rm tot}-U_1-U_2+P(V_{\rm tot}-V_1-V_2)}{T}
\label{b24}
\end{equation}
can be rather obviously translated into the minimum condition
for the convex function
\begin{eqnarray}
\tilde{G}(U_1,V_1,U_2,V_2;T,P) &=& \tilde{G}^{(1)}(U_1,V_1;T,P)+
\tilde{G}^{(2)}(U_2,V_2;T,P) \nonumber \\
&=& U_1-TS^{(1)}(U_1,V_1)+PV_1+U_2-TS^{(2)}(U_2,V_2)+PV_2\,,
\label{b25}
\end{eqnarray}
as immediately follows from computing the partial derivatives
of Eq.\,(\ref{b25}).

In closing, we re-examine the question of the shape of the
thermodynamic potentials for a system undergoing a discontinuous
phase transition. The Gibbs free energy is the thermodynamic
potential that is usually considered for describing the phases of
matter. This quantity stems from $\tilde{G}$ after tracing the
locus of its global minimum as a function of $T$ and $P$. When a
discontinuous phase transition line is approached, a piece of
ruled surface appears in the profile of the fundamental relation,
and then also in the graph of $\tilde{G}$, which remains convex,
even though not everywhere strictly convex. This implies a
discontinuous evolution for the location of the absolute
$\tilde{G}$ minimum (not for the absolute minimum itself!), {\it
i.e.}, a jump from one valley to another as soon as the
coexistence line is crossed.

While one single minimum is the rule for $\tilde{G}$ in the
thermodynamic limit, this is not generally true for a {\it finite}
system. In the framework of the statistical-mechanical foundations
of thermodynamics, this means that the microcanonical $S(U,V,N)$
of the finite system may not be everywhere
concave~\cite{Ispolatov}. In fact, near the would-be first-order
transition point, a dip will usually appear in the $S$ profile
which is responsible for the phenomenon of metastability. In turn,
$\tilde{G}$ is not everywhere convex and a competition arises
between two different local minima: while the deepest minimum
characterizes the most stable phase, the other one, as long as it
is present, will be the sign that another phase is at least
metastable (this is the usual occurrence in mean-field treatments
of first-order transitions).

The infinite-size behavior of $\tilde{G}$ is sketched in Fig.\,1,
where its typical profile close to a first-order transition point
is shown. Here, $\tilde{G}(U,V;T,P)$ is plotted as a function of
$V/N$ at constant $T$, for a number of values of $P$ across the
coexistence line relative to, say, the liquid and the vapor of a
substance (only a slice of the Gibbs surface along a locus $U(V)$
passing through the actual point of minimum of $\tilde{G}$ is
represented in the figure). The abscissa of the $\tilde{G}$
minimum gives the specific volume of the most stable system phase
for the given $T$ and $P$ values. At the same time, the $G$
profile will show, both as a function of $P$ at constant $T$ as
well as a function of $T$ at constant $P$, a cusp-like singularity
of the same kind discussed before for the Helmholtz free energy,
which is responsible for the discontinuities of volume and
entropy, respectively, across the transition.

\section{Conclusions}

Starting from the maximum entropy principle, we have provided a
detailed derivation of the principle of minimum energy and of
similar principles for thermodynamic potentials and Massieu
functions, by resorting to the mathematical theory of concave
many-variable functions. This calculation was also motivated by
the fact that standard reference books on thermodynamics usually
do not give enough information about this point. In our opinion,
proving the interchangeability of all thermodynamic
representations is a necessary prerequisite that allows the
interested reader to fully appreciate the elegance of the
thermodynamic formalism.

\vspace{3mm}
\begin{center}
{\bf Acknowledgements}
\end{center}
We thank Paolo Cubiotti for a critical reading of the manuscript.

\newpage
\begin{center}
{\bf Figure Captions} \vspace{5mm}
\end{center}
\begin{description}
\item[{\bf Fig.\,1 :}]  The figure shows how the profile of $\tilde{G}$
plotted as a function of the volume $V$ is modified when
approaching a first-order phase transition, like that from vapor
to liquid. As the pressure changes -- at constant temperature --
across the coexistence value $P_{\rm coex}$, one observes a
thermodynamic-stability crossover from the vapor to the liquid
phase (see text). Note that, with the single exception of
$P=P_{\rm coex}$, the point of minimum of $\tilde{G}$ is regular
(in particular, the tangent plane is well-defined there).
\end{description}
\newpage
%
%
\begin{figure}
\begin{center}
\setlength{\unitlength}{1cm}
\begin{picture}(16,12)(0,0)
\put(0,0){\psfig{file=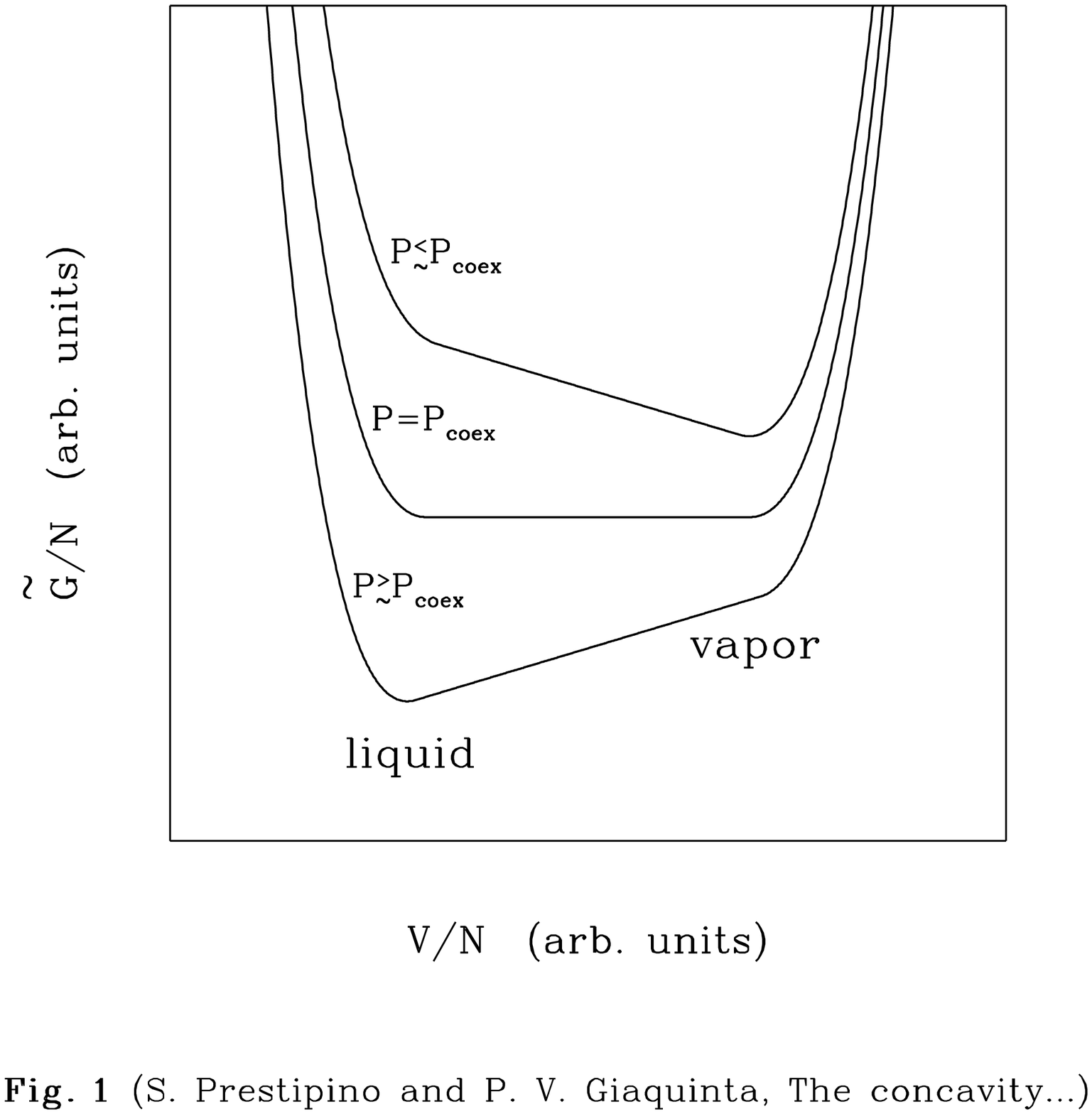,width=16cm,bbllx=12cm}}
\end{picture}
\end{center}
\end{figure}

\begin{thebibliography}{99}
\bibitem{note} Observe that for a generic quadratic form of two variables,
$H(\xi_1,\xi_2)=A\xi_1^2+2B\xi_1\xi_2+C\xi_2^2
=A\left[ \left( \xi_1+(B/A)\xi_2\right) ^2+(\xi_2^2/A^2)
\times (AC-B^2)\right] $ (where $A\neq 0$), $H$ is positive definite iff
$A>0$ and $AC-B^2>0$, and negative definite iff $A<0$ and $AC-B^2>0$.
\bibitem{note2}  Alternatively, one can define $\tilde{F}$ as $U(S,V)-TS$.
In this case, $\tilde{F}$ would be a function of $S$ which parametrically
depends on $T$ and $V$. Anyway, the minimum and the convexity properties of
$\tilde{F}$ remain the same (see Appendix B of \cite{url}).
\end{thebibliography}
\end{document}